\documentstyle[aps,preprint]{revtex}
\begin{document}
\title{Nonequilibrium phase separation in  traffic flows}
\author{Hisao Hayakawa\thanks{e-mail: hisao@phys.h.kyoto-u.ac.jp}$^1$
and Ken Nakanishi\thanks{e-mail: tmknaka@eng.shizuoka.ac.jp}$^2$}
\address{$^1$: Graduate School of Human and Environmental Studies,
Kyoto University, Kyoto 606-01, Japan \\
$^2$: Department of Mechanical Engineering, Shizuoka University,
Hamamatsu 432, Japan}
\maketitle
\begin{abstract}
Traffic jam in an optimal velocity model with the backward reference function
is analyzed. 
An analytic  scaling solution is presented 
near the critical point of  the phase separation.
The validity of the solution has been confirmed from the comparison 
with the simulation of the model.

\end{abstract}

\vspace*{0.5cm}


Recently, the importance  of cooperative behavior in
 dissipative systems 
consisting of discrete elements
has been recognized among physists.
As a result,  granular materials have been studied extensively from
physical point of views.\cite{granules}. 
Similarly, to know the properties of
 traffic jams in daily life is also  an attractive subject 
 not only for engineers but also physists\cite{traffic}.
There are some similarities between two phenomena in particular 
in the simplest situation where
 cars and particles are respectively confined in  
a highway and a long tube. 
Thus,  it is interesting to clarify  common and universal 
mathematical structure behind these phenomena.

We propose here a model of the traffic flow
\begin{equation}\label{rOV}
 \ddot x_n=a[U(x_{n+1}-x_n)V(x_{n}-x_{n-1})-\dot x_n],
\end{equation} 
where $x_n$ and $a$ are the positions of $n$ th car, and 
the sensitivity, respectively. 
This model contains  the psychological effect of drivers.
Namely,  the driver of $x_n$ takes care of not only the distance
ahead $x_{n+1}-x_n$ but also the backward distance 
 $x_n-x_{n-1}$. 
The optimal velocity function $U$ should be a monotonic increasing 
function of the distance of $x_{n+1}-x_n$ and $V-1$
 should be a monotonic decreasing function
of $x_n-x_{n-1}$. Thus,  we adopt
\begin{equation}\label{back}
U(h)=\tanh(h-2)+\tanh(2); \quad  V(h)=1+f_0(1-\tanh(h-2))
\end{equation}
for the later explicit calculation.
We put these optimal velocity 
functions as the product form $UV$ in (\ref{rOV}),
 because the driver of $x_n$ cannot accelerate the car without enough
the forward distance $x_{n+1}-x_{n}$ even when the distance $x_n-x_{n-1}$
becomes short.
This model (\ref{rOV}) with (\ref{back}) is the generalization of
the optimal velocity (OV) model proposed by Bando et al. \cite{bando}
\begin{equation}\label{OV}
\ddot x_n=a[U(x_{n+1}-x_n)-\dot x_n].
\end{equation} 
Our model is also similar to the model of granular flow in a one dimensional 
tube
\begin{equation}\label{powder}
\ddot x_n=\zeta[\tilde U(x_{n+1}-x_{n-1})-\dot x_n]+g(x_{n+1}-x_n)
-g(x_n-x_{n-1})
\end{equation}
where the explicit forms of $\tilde U$ and the force $g$ are not
important in out argument.
Although real systems contain variety of  
cars and particles and higher dimensional
effects, we believe that 
the most essential parts of both traffic flows and granular 
 pipe flows can be  understood by pure one dimensional models
(\ref{rOV}) and (\ref{powder}).
The reason is as follows:
 It is known\cite{chuo} that
 model (\ref{powder}) supplemented by the white noise produces
a power law in the frequency spectrum of the density correlation function
$S(q,\omega)\sim \omega^{-4/3}$,
 whose exponent 4/3 is very close to
the experimental value\cite{chuo} and that by the lattice-gas 
automata simulation\cite{peng}. 
From this success 
 the essential effects of randomness
such as passing cars and variety of cars
seem to be represented by the adding white noise to the models 
(\ref{rOV}) and (\ref{powder}).

Komatsu and Sasa\cite{komatsu} reveal that the original
 OV model 
can be 
reduced into the modified Korteweg-de Vries (MKdV) equation 
at the critical point (the averaged car distance $h=2$)
 for the phase separation. 
They also show that   
symmetric kink solitons deformed by
dissipative corrections describe a bistable phase separation. 
The exactly solvable models in which the essential 
characteristics of the optimal velocity model
are included have been proposed\cite{exact}. 
 However, as will be shown, 
the generalized optimal  velocity model (\ref{rOV}) and 
granular model (\ref{powder}) as well as
 the fluid model of traffic flows by Kerner and Konh\"auser\cite{KK}
and two fluid models in granular flows\cite{fluid}  
are not reduced to MKdV equation   but 
exhibit the phase separations 
between a linearly unstable phase and a stable phase\cite{komatsu2}. 
Thus, there is a wider universality class of dissipative particle dynamics
which contains (\ref{rOV}), (\ref{powder}) and fluid models\cite{KK,fluid}.

The aim of this Letter is to obtain an analytic scaled solution
of (\ref{rOV}).
 To demonstrate quantitative validity
of our analysis we will compare it with the result of our simulation.
After the completion of our analysis on (\ref{rOV}), we will briefly
discuss the relation of the result and 
the expected results in  (\ref{powder}) and fluid models.


Let us rewrite (\ref{rOV})  as
\begin{equation}\label{rOV2}
\ddot r_n=a[U(h+r_{n+1})V(h+r_n)-U(h+r_n)V(h+r_{n-1})-\dot r_n]
\end{equation}
where   $h$ is the averaged distance of successive cars
 and $r_n$ is $x_{n+1}-x_n-h$ .

Now, let us consider the linear stability of (\ref{rOV2}).  
The linearized equation of (\ref{rOV2}) around $r_n(t)=0$
 is given by
\begin{equation}\label{linear} 
\ddot  r_n=a[U'(h)V(h)( r_{n+1}- r_n)
+U(h)V'(h)( r_n- r_{n-1})-\dot  r_n]
\end{equation}
where  the prime refers to the differentiation 
with respect to the argument. 
 With the aid of the Fourier transformation
$r_q(t)=\frac{1}{N}\sum_{n=1}^N \exp[-i q n h]  r_n(t)$
with $q=2\pi m/N h$ and the total number of cars $N$ 
we can rewrite (\ref{linear}) as
\begin{equation}\label{linearF}
(\partial_t-\sigma_+(q))(\partial_t-\sigma_-(q))r_q(t)=0
\end{equation}
with
\begin{equation}\label{sigmapm}
\sigma_{\pm}(q)=-\frac{a}{2} 
\pm \sqrt{ (a/2)^2 -a D_h[U,V]( 1-\cos(q h) )+
i a (UV)'\sin(q h)} ,
\end{equation} 
where  we drop the argument $h$ in $U$ and $V$.
 $D_h[U,V]\equiv U'(h)V(h)-U(h)V'(h)$ denotes Hirota's derivative.
The solution of the initial value problem in 
(\ref{linearF}) is the linear combination of terms in proportion to 
$\exp[\sigma_+(q) t]$ and $\exp[\sigma_-(q) t]$.
The mode in proportion to  $\exp[\sigma_-(q) t]$
can be interpreted as  the fast decaying mode, while the term in 
proportion to   $\exp[\sigma_+(q) t]$ is the slow and more important  mode.

The violation of the linear
stability of the uniform solution in (\ref{linear}) is equivalent to 
$Re[\sigma_+(q)]\ge 0$. Assuming $qh\ne 0$ ($qh=0$ is the neutral mode),
the instability condition is given by
$2 (UV)'^2\cos^2(\frac{q h}{2})\ge a D_h[U,V].
$
Thus, the most unstable mode exists at $q h\to 0$ and the neutral curve for 
long wave instability is given by
\begin{equation}\label{neutral}
a=a_n(h)\equiv \frac{2(UV)'^2}{D_h[U,V]} .
\end{equation}
The neutral curve 
in the parameter space $(a,h)$ is shown in Fig.1 
for  $f_0=1/(1+\tanh(2))$ in (\ref{back}). 
For later convenience, we write the explicit form of 
the long wave expansion of $\sigma_+$
in the vicinity of the neutral line 
\begin{equation}
\sigma_+(q)= 
i c_0q h-c_0^2\frac{a-a_n( h)}{a_n( h)^2}(qh)^2
-i\frac{(q h)^3}{6}c_0-
\frac{(qh)^4}{4a_n( h)}c_0^2+O((qh)^5)
\label{dispersion}
\end{equation}
where $c_0=(UV)'$.
Thus, the uniform state becomes 
unstable due to the negative diffusion constant appears for $a<a_n(h)$.


The simplest way to describe hydrodynamic mode is the long wave expansion.
It is easy to derive the KdV equation near the neutral curve 
from (\ref{rOV}) as in the case of fluid models\cite{KK,fluid}.
 To describe the phase separations, however,  we should  choose
the critical point $(a,h)=(a_c,h_c)$ from the cross point of 
$(U(h)V(h))''=0$ where the coefficient of $\partial_x r^2$ becomes 
zero and the neutral curve, because KdV equation only has pulses
while cubic nonlinear terms can produce the interface solution
to connect two separated domains.
 The explicit critical point of (\ref{back}) with
$f_0=1/(1+\tanh(2))$  is given by
\begin{equation}\label{cp}
h_c= 2-\tanh^{-1}(1/3)\simeq 1.65343; \quad 
a_c=\frac{512}{81}f_0^2\simeq 1.63866. 
\end{equation}
Unfortunately, the reduced equation based 
on the long wave expansion of 
our model is an ill-posed equation. In fact, 
 the scalings 
of variables as $ r_n(t)=\epsilon
 r(z,\tau)$
,$ z=\epsilon
(x+c_0t)$ and $\tau=\epsilon^3 t$ with $\epsilon=\sqrt{(a_c-a)/a_c}$
leads to 
\begin{equation}\label{long}
\partial_{\tau}r=a_1\partial_z r^3-a_2\partial_z^3r+a_3
\partial_z^2r^2
\end{equation}
in the lowest order, where
  $a_1$, $a_2$ and $a_3$ are constants.
Its linearized equation
around $r=d_0$ is unstable for all scale, because
the solution with $r-d_0\simeq \exp[ikz+\lambda_k \tau]$ 
has the growth rate 
 $Re[\lambda_k]=2  k^2 a_3 d_0$ which is always positive when $a_3 d_0>0$.

Of course, this irregularity in the short scale is from the long wave
approximation. The regularity of the original model (\ref{rOV}) 
can be checked easily as
follows: Let $x=n h$ be regarded as a continuous variable.
From $r(x\pm h, t)=\exp[\pm h \partial_x]r(x,t)$
or the  Fourier component $\exp[\pm i q h]$
of the translational operator $ \exp[\pm h \partial_x]$,
the function of the translational operation in the shortest scale ($qh=\pi$
in the Fourier space) is
$r(x\pm h,t)\to -r(x,t)$. Thus, our model in (\ref{rOV2}) 
for $r(x,t)$ in the shortest scale 
is reduced to 
\begin{equation}\label{short}
\partial_t^2 r=a[W(r)-\partial_t r], \quad
W(r)\equiv U(h-r)V(h+r)-U(h+r)V(h-r) .
\end{equation}   
Substituting (\ref{back}) into (\ref{short}) it is easy to show  
$W'(r)=-({\rm sech}^2(h-r-2)+{\rm sech}^2(h+r-2))
\{1+f_0(1+\tanh(2))\}<0$.
Then, the growth rate of 
the linearized equation of (\ref{short}) by
 $r-d_0\sim \exp[\lambda t
+i k x]$ is given by 
\begin{equation}
\lambda=\frac{-a\pm \sqrt{a^2-4a|W'(d_0)|}}{2},
\end{equation}
where $Re[\lambda]\le 0$ for any $d_0$.  Thus, 
the original model (\ref{rOV}) is stable for the perturbation in
the short scale.

Although it is possible to derive a regularized long wave equation thanks to
the Pade approximation\cite{ooshida},
the result is more complicated than the original model (\ref{rOV}).
Thus, to obtain the asymmetric 
propagating kink solution,
  we only eliminate the fast decaying mode
in (\ref{rOV2}) 
as
\begin{equation}\label{lattice}
(\partial_t -\sigma_+(\partial_x))r(x,t)=(\sigma_+-\sigma_-)^{-1}N[r(x,t)] ,
\end{equation}
where $N[r]$ represents the nonlinear terms coming from $UV$.
Since $(\sigma_+-\sigma_-)^{-1}$ is the inverse of the polynomial of 
the differential operators, it is convenient to use the 
expansion $(\sigma_+-\sigma_-)^{-1}\simeq 
a^{-1}[1-\displaystyle\frac{2 h}{a}(UV)'\partial_x
+O(h^2)]$.
Equation (\ref{lattice}) is the regularized partial differential equation.

To obtain the scaled propagating kink solution 
we assume the scaling of the variables by $\epsilon=\sqrt{(a_c-a)/a_c}$ as
\begin{equation}\label{scaling}
r(x,t)=\epsilon \displaystyle\sqrt{\frac{6 c}{|(UV)'''|}}
R(z)
,\quad z=\epsilon \displaystyle\sqrt{\frac{6 c}{c_0}}
(\frac{x}{h}+c_0t-\epsilon^2 c t)
\end{equation}
where the argument is fixed at $h=h_c$,
 and $c$ is the positive free parameter which will be determined 
from the perturbation analysis.
Substituting (\ref{scaling})
into (\ref{lattice}) and 
use the expansion 
\begin{equation}\label{N[r]}
N[r]/a= \sum_{n=1}^{\infty}\sum_{m=2}^{\infty} h^mC_{mn} \partial_x^n r^m
- h^3 U'V'\partial_x r\partial_x^2 r+\cdots
\end{equation}
where $C_{21}=\frac{1}{2}(UV)'',  C_{22}=\frac{1}{4}D_h[U,V]', 
C_{23}=\frac{1}{12} (UV)'' , 
C_{31}=\frac{1}{6}(UV)''',  C_{32}=\frac{1}{12}D_h[U,V]'', 
C_{41}=\frac{1}{24}(UV)''''$, 
 and integrate by part of (\ref{lattice}),  we obtain
\begin{equation}\label{propagate}
\frac{d^2R}{dz^2}- R(R^2-1)+\beta \frac{d}{dz}(R^2)=
\epsilon\displaystyle\sqrt{\frac{c}{c_0}}
M[R], 
\end{equation}
where $\beta=3D_h[U,V]'/(2\sqrt{c_0|(UV)'''|})$.
Here we neglect the contribution from the boundary and   
\begin{equation}\label{M}
M[R]=\rho_{23}\left(\frac{d R}{dz}\right)^2-
\rho_{32}\frac{dR^3}{dz}-\rho_{41}R^4
-\frac{1}{4\eta}(4 \frac{dR}{dz}+\frac{d^3R}{dz^3}-
2\frac{c_0}{c}\frac{dR}{dz}) ,
\end{equation}
where $1/\eta=\sqrt{6}D_h[U,V]/c_0$,
, $\rho_{23}=3 \sqrt{6}U'V'/\sqrt{c_0|(UV)'''|}$,
$\rho_{32}=\sqrt{3/2}D_h[U,V]''/|(UV)'''|$ and 
$\rho_{41}=\sqrt{3c_0}(UV)''''/(2\sqrt{2|(UV)'''|^3})$.
Assuming $R(z)=R_0(z)+\epsilon R_1(z)+\cdots$,
 we obtain an asymmetric kink-antikink
solution
\begin{equation}\label{asymmetry}
R_0(z)=\tanh(\theta_{\pm}z); \quad \theta_{\pm}=
\frac{\beta\pm \sqrt{\beta^2+2}}{2}
\end{equation}
in the lowest order.
The linearized equation of (\ref{asymmetry}) can be reduced to
\begin{equation}\label{linearized}
{\cal L} R_1=\displaystyle\sqrt{\frac{c}{c_0}}
 M[R_0];\quad {\cal L}= \frac{d^2}{dz^2}
+1-3 R_0^2+2\beta (R_0\frac{d}{dz}+\frac{d R_0}{dz}).
\end{equation}
The solvability condition to determine $c$ is 
\begin{equation}\label{solva}
(\Phi_0,M[R_0])\equiv \int_{-\infty}^{\infty} dz \Phi_0 M[R_0]=0,
\end{equation}
where $\Phi_0$ satisfies
${\cal L}^{\dagger}\Phi_0=0$.
The explicit form of $\Phi_0$ is given by
\begin{equation}\label{zero-fun}
\Phi_0(z)=({\rm sech}[\theta_{\pm} z])^{1/\theta_{\pm}^2}.
\end{equation}
Thus, the solvability condition is reduced to 
\begin{equation}\label{determine}
\frac{c_0}{c}=
2+\theta_{\pm}^2\left(2-3\frac{I_2^{(\pm)}}{I_1^{(\pm)}}\right)
+2\eta[ 3\rho_{32}\left(1-\frac{I_2^{(\pm)}}{I_1^{(\pm)}}\right)
+\frac{\rho_{41}}{\theta_{\pm}}
\left(\frac{I_0^{(\pm)}}{I_1^{(\pm)}}-2
+\frac{I_2^{(\pm)}}{I_1^{(\pm)}}\right)-
\rho_{23}\theta_{\pm}\frac{I_2^{(\pm)}}{I_1^{(\pm)}}]
\end{equation}
where $I_n^{(\pm)}=\int_{-\infty}^{\infty} dx ({\rm sech} x)^{1/\theta_{\pm}^2+2n}
=\sqrt{\pi}\displaystyle\frac{\Gamma(1/(2\theta_{\pm}^2)+n)}
{\Gamma(1/(2\theta_{\pm}^2)+n+1/2)}$.

To obtain the explicit form we adopt $f_0=1/(1+\tanh(2))$ in (\ref{back}).
In this case the coefficients in (\ref{determine}) are reduced to 
$\rho_{23}=-3/2$, $\rho_{32}=-\beta$, $\rho_{41}=-1/4$, $\eta=1/(4\beta)$
$c_0=2^6 f_0/3^3=1.20689$,  
and $\beta=3\sqrt{3}/(8\sqrt{2}f_0)=
0.902037$. 
Thus, we obtain $c$ as
\begin{equation}\label{selection}
c_+= 0.62485945 ;\quad c_-=0.82170040 ,
\end{equation}
where $c_{\pm}$ are respectively  the solution of
(\ref{determine})  corresponding to  $\theta_{\pm}$.
Since there are two propagating velocities, the linearly stable region
invades the unstable region if there are many domains in the system.

To check the validity of our analysis we perform the numerical simulation
of (\ref{rOV}) and (\ref{back}) with $f_0=1/(1+\tanh(2))$
under the periodic boundary condition. We adopt 
the classical fourth-order Runge-Kutta scheme. Since our purpose is 
the quantitative test of (\ref{asymmetry}) and (\ref{selection}),
the initial condition is restricted to the localized symmetric
form $r_n= 18.7/N(\tanh(n-N/4)-\tanh(n-3N/4)-1)$ where $N$ is
the number of cars. Taking into account the scaling properties
we perform the simulation for  the set of parameters $(\epsilon, N)=
(1/2, 32), (1/4,64), (1/8,128), (1/16,256)$  until $r_n$ relaxes to
 steady propagating states. Our  result is plotted in Figs.1 and 2.
Figure 1 displays  points which have the maximum and the minimum values
of successive car distance in
each parameter set, and theoretical coexistence curve 
\begin{equation}\label{coex}
a=a_c\left(1- \frac{(h-h_c)^2}{A^2}\right); 
\quad A\equiv \displaystyle\sqrt{\frac{6\bar c}{|(UV)'''|}}
=1.15850495992,
\end{equation} 
where the agreement with each other is obvious.
Notice we  adopt $\bar c=(c_++c_-)/2$ as the traveling velocity,
because the domain cannot move due to the finite size effect. 
From this figure we can see that 
one of the branches is in the linearly unstable region but
the theoretical curve  recovers the simulation result.    
Figure 2 demonstrates that the numerical result has a scaling solution
which has an asymmetric kink-antikink pair.
The linear combination of  our theoretical curve
(\ref{asymmetry}) and (\ref{selection}) is plotted as the solid line
by choosing the position of the kink and the antikink. Our theoretical 
curve  agrees with simulation value without other fitting parameters.
The quantitative discussion on the spreading process due to 
the finite $c_+-c_-$
will be discussed elsewhere.

Let us comment on the universality class of traffic flows and 
granular pipe flows. All of models introduced here except for (\ref{OV}) have
asymmetric kink-antikink pairs and qualitatively resemble 
behaviors with each other.
Komatsu \cite{komatsu2} has derived 
(\ref{long}) as the long wave equation from the fluid model of
traffic flow\cite{KK}  which is equivalent to the two-fluid models
in granular flows. It is also easy to  derive (\ref{long}) 
from (\ref{powder}). In this sense,
granular flows and traffic flows compose a universality class 
and our discussion here  essentially can be used in any models
for traffic flows and granular pipe flows.
On the other hand, OV model in (\ref{OV}) is a special case of
the above generalized models. For example, the models 
with $f_0=0$ in (\ref{back}) 
and $g''(a_c)=0$ in (\ref{powder}) are reduced  to MKdV equation. 

In conclusion, we obtain the analytic scaling solution of (\ref{rOV}) and
(\ref{back}) which describes the phase separation between 
a linearly unstable phase and a stable phase. The accuracy 
and relevancy of the solution are confirmed by the direct simulation.

One of the authors (HH) thanks Ooshida, T. for fruitful discussion.
This work is partially supported by Grant-in-Aid of Ministry of Education,
Science and Culture of Japan (09740314).

\begin{figure}
\caption{
Plots of the coexistence curve (26), the solid line, and
the neutral curve (9), the dashed line, as the functions of
$h$. The scattered points are the maximum and the minimum  
distances of successive cars
for a given $a$ obtained from our simulation.
}
\label{fig1}
\end{figure}

\begin{figure}
\caption{
The linear combination of our theoretical curve (20)
 and the scaled simulation data of the relative distance of succesive cars for
$(\epsilon,N)=(1/2,32),(1/4,64),(1/8,128),(1/16,256)$,
where $'N.s'$ denotes the scaled data for $N$ cars systems.
The solid curve is 
$f(z)=\tanh(\xi \theta_+(z-z_+))-1+\tanh(\xi\theta_-(z-z_-))$
with $\xi=(6 \bar c/c_0)^{1/2}/16=0.11851533$ and
two fitting parameters $z_+=62.5 $ and $z_-=190.5$,
 where the spatial scale is 
measured by the average distance in $N=256$.  
}
\label{fig2}
\end{figure}

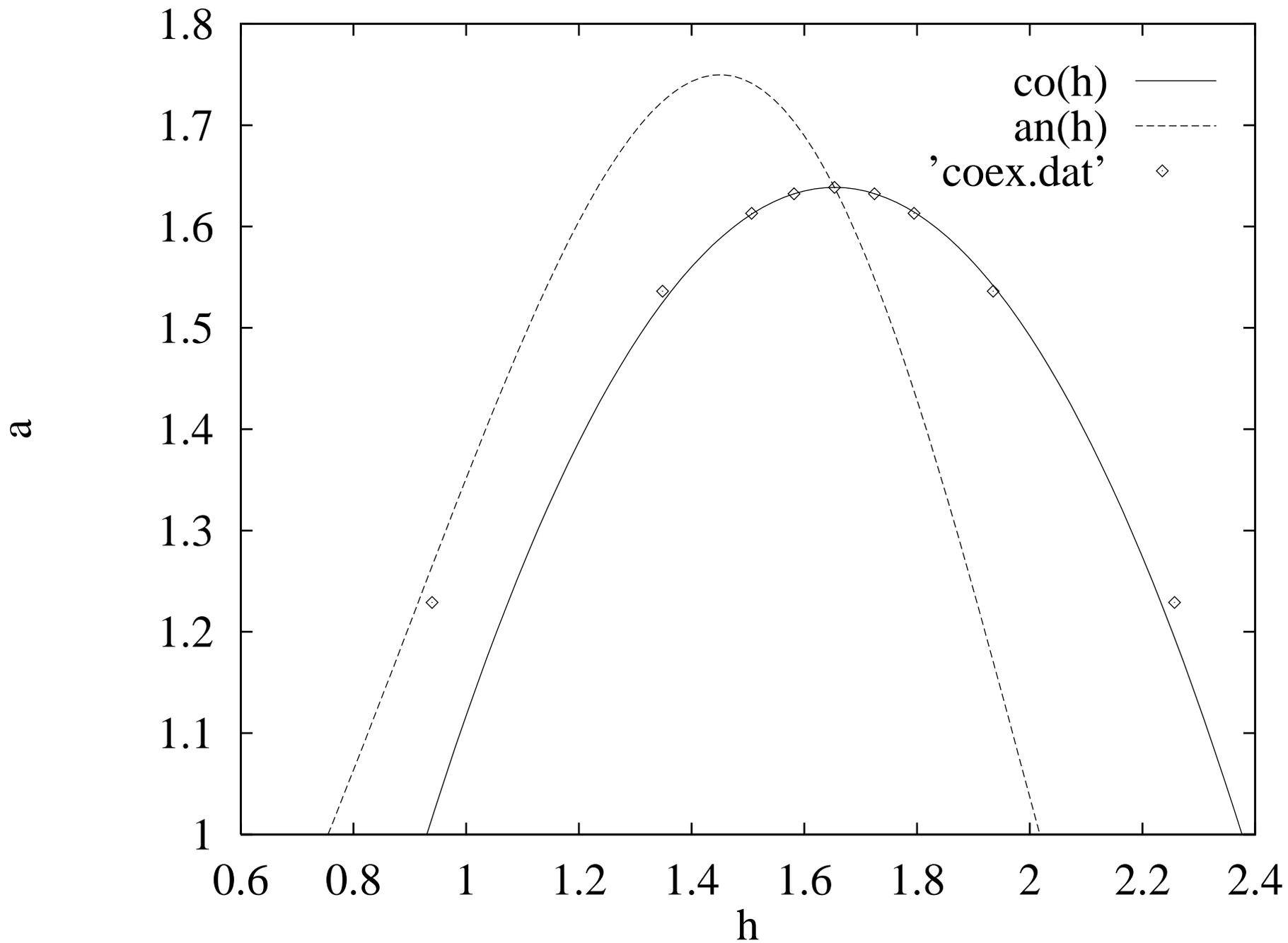
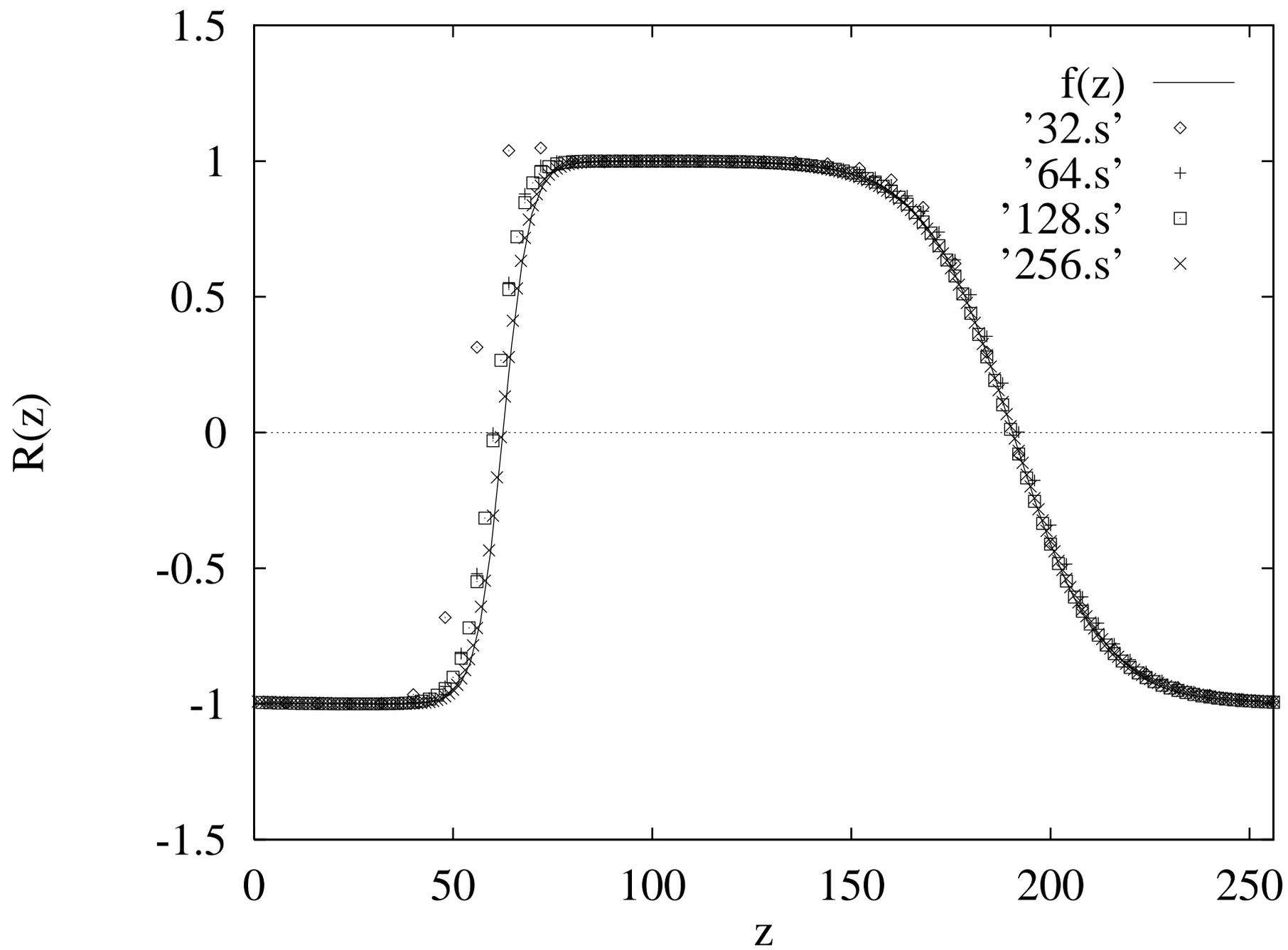
\end{document}